\documentclass[twocolumn, switch]{article} 

\usepackage{preprint}

\usepackage{amsmath, amsthm, amssymb, amsfonts}
\usepackage{bm}
\usepackage{amsmath}
\usepackage{graphicx}

\usepackage{setspace}
\usepackage{booktabs}
\usepackage{array}
\usepackage{color}
\usepackage{multirow}
\usepackage{multicol}
\usepackage{url}
\usepackage{cite}
\usepackage{amsmath,amssymb,amsfonts}
\usepackage{algorithmic}
\usepackage{graphicx}
\usepackage{bm}
\usepackage{array}
\usepackage{bbding}
\usepackage{multirow}
\usepackage[flushleft]{threeparttable}
\usepackage{amssymb}
\usepackage{pifont} 
\usepackage{booktabs}
\usepackage{textcomp}
\usepackage{xcolor}
\usepackage{tabularx}
\usepackage{makecell}
\usepackage{caption}
\usepackage{xcolor}
\usepackage[page]{appendix}

\usepackage[numbers,square]{natbib}
\usepackage[utf8]{inputenc}	
\usepackage[T1]{fontenc}	
\usepackage{xcolor}		
\usepackage[colorlinks = true,
            linkcolor = blue,
            urlcolor  = blue,
            citecolor = blue,
            anchorcolor = black]{hyperref}	
\usepackage{booktabs} 		
\usepackage{nicefrac}		
\usepackage{microtype}		
\usepackage{lineno}		
\usepackage{float}			
\usepackage{multicol}		

\definecolor{myblue}{HTML}{6c8ebf}
\definecolor{myred}{HTML}{b85450}

\usepackage{newfloat}
\DeclareFloatingEnvironment[name={Supplementary Figure}]{suppfigure}
\usepackage{sidecap}
\sidecaptionvpos{figure}{c}

\usepackage{titlesec}
\titlespacing\section{0pt}{12pt plus 3pt minus 3pt}{1pt plus 1pt minus 1pt}
\titlespacing\subsection{0pt}{10pt plus 3pt minus 3pt}{1pt plus 1pt minus 1pt}
\titlespacing\subsubsection{0pt}{8pt plus 3pt minus 3pt}{1pt plus 1pt minus 1pt}

\usepackage{graphics}
\usepackage{color}
\usepackage{hhline}
\usepackage{lipsum}

\usepackage{hyperref}
\usepackage{bm}
\usepackage[flushleft]{threeparttable}
\usepackage{bbding}
\usepackage{float}
\usepackage{subfig}
\usepackage{multirow}
\usepackage[inline]{enumitem}
\usepackage[ruled]{algorithm2e}

\title{DiffVC-OSD: One-Step Diffusion-based Perceptual Neural Video Compression Framework}

\usepackage{authblk}

\author[ ]{Wenzhuo Ma}
\author[*]{Zhenzhong Chen}
\affil[ ]{School of Remote Sensing and Information Engineering, Wuhan University}

\begin{document}

\twocolumn[ 
  \begin{@twocolumnfalse} 
  
\maketitle

\begin{abstract}

In this work, we first propose DiffVC-OSD, a One-Step Diffusion-based Perceptual Neural Video Compression framework. Unlike conventional multi-step diffusion-based methods, DiffVC-OSD feeds the reconstructed latent representation directly into a One-Step Diffusion Model, enhancing perceptual quality through a single diffusion step guided by both temporal context and the latent itself. To better leverage temporal dependencies, we design a Temporal Context Adapter that encodes conditional inputs into multi-level features, offering more fine-grained guidance for the Denoising Unet. Additionally, we employ an End-to-End Finetuning strategy to improve overall compression performance. Extensive experiments demonstrate that DiffVC-OSD achieves state-of-the-art perceptual compression performance, offers about 20$\times$ faster decoding and a 86.92\% bitrate reduction compared to the corresponding multi-step diffusion-based variant.

\end{abstract}

\vspace{0.4cm}

  \end{@twocolumnfalse} 
] 

\newcommand\blfootnote[1]{%
\begingroup
\renewcommand\thefootnote{}\footnote{#1}%
\addtocounter{footnote}{-1}%
\endgroup
}

\section{INTRODUCTION}
\vspace{-15pt}
{\blfootnote{Corresponding author: Zhenzhong Chen, E-mail:zzchen@ieee.org}}

Neural Video Compression (NVC) has progressed rapidly, surpassing traditional codecs in pixel-level distortion metrics. However, due to the inherent trade-off among rate, distortion, and perception \cite{rate_distortion_perception}, distortion-focused methods often yield suboptimal perceptual quality. This has led to growing interest in perceptual NVC, which emphasizes the rate-perception trade-off to produce clearer, more detailed, and realistic reconstructions. While GAN-based methods enhance detail through adversarial training, they remain limited by unstable training dynamics. Recently, diffusion models pre-trained on large-scale image-text datasets have emerged as a promising paradigm for perceptual NVC.

Existing diffusion-based NVCs typically adopt multi-step diffusion models, which reconstruct high perceptual quality results through multiple denoising steps. Liu et al. \cite{I2VC} proposed I$^2$VC, a unified multi-step diffusion-based NVC framework supporting detail-rich reconstructions under AI, LD, and RA modes. Ma et al. \cite{DiffVC} introduced DiffVC, which integrates a foundational diffusion model into the conditional coding paradigm using context from previously decoded frames to guide high-quality reconstruction. Although the above multi-step diffusion-based NVCs have demonstrated remarkable potential in perceptual quality, they still suffer from three major limitations. First, multi-step diffusion models usually start denoising from pure noise or noisy latent representations, which discard important structural information embedded in the latent features—information that is crucial for the compression task. Second, due to the iterative nature of multi-step inference, these models suffer from high inference latency, especially in video compression tasks where each frame requires multiple diffusion steps. Finally, due to the heavy computational burden from gradient accumulation, multi-step diffusion-based NVCs are difficult to optimize in an end-to-end manner, which limits their overall compression performance.

However, the challenges faced by the aforementioned multi-step diffusion models can be effectively addressed by a one-step diffusion model. To this end, we propose \textbf{DiffVC-OSD} (\textbf{O}ne-\textbf{S}tep \textbf{Diff}usion-based Perceptual Neural \textbf{V}ideo \textbf{C}ompression), the first framework to introduce a One-Step Diffusion Model into NVC. Specifically, instead of adding noise to the reconstructed latent representation, we directly feed it into the diffusion model. By leveraging the structural information preserved in the latent and the powerful generative capacity of the diffusion model, high perceptual quality reconstructions can be achieved in just a single diffusion step. Furthermore, to more effectively integrate the one-step diffusion model into the conditional coding paradigm, we introduce a Temporal Context Adapter (TCA), which encodes the temporal context extracted from the previously decoded frame and the reconstructed latent representation into multi-level features to guide the Denoising Unet in a fine-grained manner. Finally, we employ an End-to-End Finetuning strategy that jointly optimizes the one-step diffusion model and other modules for the rate-distortion-perception trade-off, enabling optimal compression performance. Extensive experiments demonstrate that DiffVC-OSD achieves state-of-the-art perceptual compression performance across all test sets, while providing a 86.92\% bitrate reduction and 20× faster decoding compared to its multi-step diffusion-based counterpart.

\begin{figure*}[!t]
	\centering
	\includegraphics[width=1.0\textwidth]{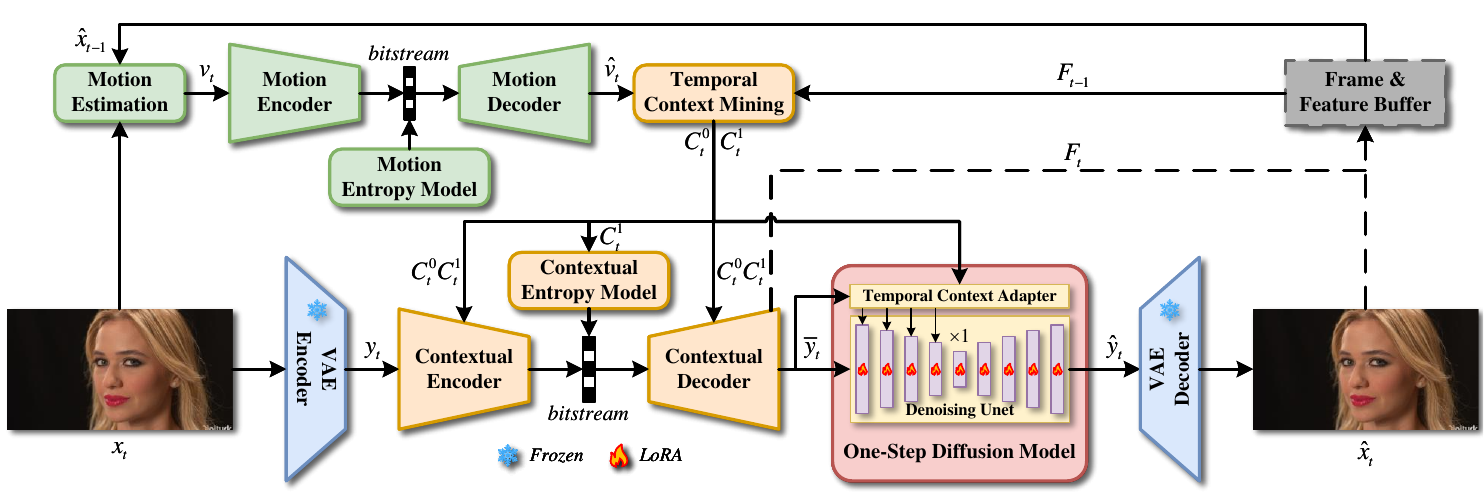}
	\caption{The framework of DiffVC-OSD.}
	\label{fig:DiffVC-OSD}
	\vspace{-15pt}
\end{figure*}

\vspace{-10pt}

\section{Methodology} \label{sec:method}
As illustrated in Figure \ref{fig:DiffVC-OSD}, we propose a One-Step Diffusion-based Perceptual Neural Video Compression framework, termed DiffVC-OSD. Built upon DiffVC \cite{DiffVC}, DiffVC-OSD follows the conditional coding paradigm and leverages Stable Diffusion V2.1-base \cite{SD} as its foundational diffusion model. The overall pipeline of DiffVC-OSD is as follows: the Motion Modules (highlighted in green) first estimate and encode the motion vector $v_t$ between the current frame $x_t$ and the previously decoded frame $\hat{x}_{t-1}$. Next, $x_t$ is encoded into a latent representation $y_t$ using the encoder of Stable Diffusion. The Contextual Modules (highlighted in orange) utilize the feature $F_{t-1}$ of the previously decoded frame along with the reconstructed motion vector $\hat{v}_t$ to perform motion compensation, and extract multi-level temporal contexts ($C_t^0$ and $C_t^1$) to support the encoding and decoding of $y_t$. The One-Step Diffusion Model (highlighted in red) takes the noise-free reconstructed latent representation $\bar{y}_t$ as input and refines it in a single diffusion step, using both the large-scale temporal context $C_t^0$ and $\bar{y}_t$ itself as conditions to enhance perceptual quality. Finally, the enhanced latent $\hat{y}_t$ is decoded by the Stable Diffusion decoder to produce the reconstructed frame  $\hat{x}_t$ with rich details and a realistic appearance. In the following sections, we present detailed descriptions of the proposed One-Step Diffusion Model, Temporal Context Adapter, and End-to-End Finetuning strategy.
\vspace{-10pt}

\subsection{One-Step Diffusion Model} \label{sec:OSD}
In multi-step diffusion-based NVC, the reconstructed latent representation $\bar{y}_t$ is typically perturbed by Gaussian noise until it approximates pure noise, and then a denoising process is performed through multiple diffusion steps (e.g., 50 steps) to obtain the enhanced latent $\hat{y}_t$. However, the noisy latent loses important structural information that is crucial for compression task, and the need for multiple diffusion steps per video frame leads to significant inference latency, severely limiting the potential of multi-step diffusion-based NVC.

To address these issues, we introduce the One-Step Diffusion Model. Specifically, instead of adding noise, we directly feed $\bar{y}_t$ into the Denoising Unet, thereby preserving the structural information embedded in $\bar{y}_t$ and providing a more informative starting point for the diffusion process. Considering the temporal dependencies in video and the 'ready-made' temporal context in conditional coding paradigms, we concatenate $\bar{y}_t$ with large-scale temporal context $C_t^0$ as conditions $c_t$ to guide the diffusion model to generate a perceptually enhanced $\hat{y}_t$. Following the DDPM \cite{DDPM}, the one-step diffusion process can be described as:
\begin{equation}
	\begin{aligned}
		\epsilon_{\theta} &=Unet(\bar{y}_t, c_t, n), where\ c_t = Concat(\bar{y}_t, C_t^0) \\
		\hat{y}_t &= \frac{1}{\sqrt{\alpha^n}}(\bar{y}_t-\frac{1-\alpha^n}{\sqrt{1-\bar{\alpha}^n}}\epsilon_{\theta})
	\end{aligned}
\end{equation}
where $\epsilon_{\theta}$, $n$ and  $\alpha^n$/$\bar{\alpha}^n$ denote the noise predicted by the Denoising Unet, the input timestep to the Unet, and the noise scheduler coefficient of the forward diffusion process at timestep $n$ , respectively. Leveraging the powerful generative prior of the foundational diffusion model, the One-Step Diffusion Model is capable of significantly enhancing perceptual quality with only a single diffusion step.

\subsection{Temporal Context Adapter} \label{sec:TCA}
\begin{figure}[!t]
	\centering
	\includegraphics[width=0.5\textwidth]{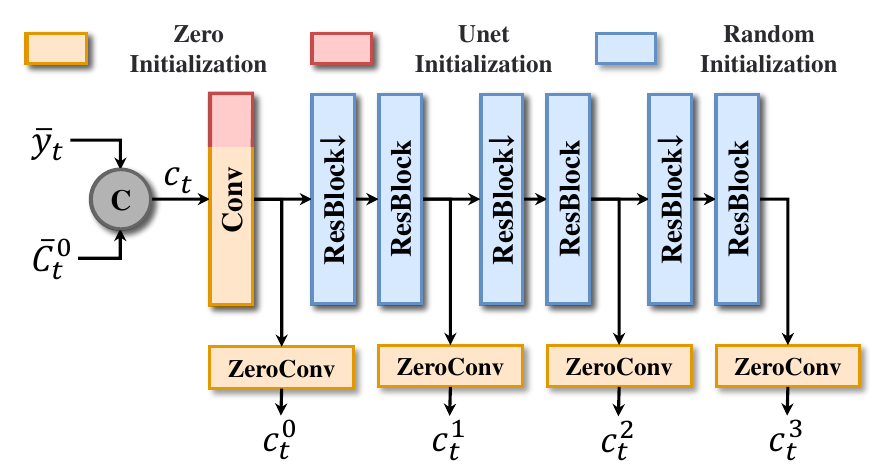}
	\caption{The structure of TCA.}
	\label{fig:TCA}
	\vspace{-18pt}
\end{figure}

To enable the Denoising Unet to better leverage the condition $c_t$ by incorporating temporal context from the previously decoded frame—rather than relying solely on the current frame—we propose the Temporal Context Adapter (TCA), as illustrated in Figure \ref{fig:TCA}. Specifically, TCA takes the concatenation of $\bar{y}_t$ and $\bar{C}_t^0$ as input and extracts multi-level features through convolutional layers and ResBlocks.
Inspired by ControlNet \cite{ControlNet}, we adopt a special initialization strategy for the first convolutional layer of TCA: the first $N_{\bar{y}_t}$ channels are initialized with the weights from the first convolutional layer of the Unet, while the remaining $N_{\bar{C}_t^0}$ channels are initialized to zero. Here, $N_{\bar{y}_t}$ and $N_{\bar{C}_t^0}$ denote the number of channels in $\bar{y}_t$ and $\bar{C}_t^0$, respectively.
Furthermore, each intermediate feature produced by TCA is passed through a zero-initialized convolutional layer to generate the final multi-level features $\{c_t^0, c_t^1, c_t^2, c_t^3\}$. These special designs allows TCA to gradually and stably learn to extract useful features, thereby facilitating the Unet in generating $\hat{y}_t$ with high perceptual quality.

\vspace{-10pt}

\begin{table*}[t]
	\centering
	\footnotesize
	\renewcommand {\arraystretch}{1.05}
	\caption{BD-rate$\downarrow$ (\%) / BD-metric$\uparrow$ for different methods on HEVC, MCL-JCV and UVG dataset. The anchor is VTM-17.0.}
	\label{tab:results}
	\resizebox{0.9\textwidth}{!}{
		\begin{threeparttable}
			\begin{tabularx}{\textwidth}{>{\raggedright\arraybackslash}p{0.7cm} >{\raggedright\arraybackslash}p{2.25cm} *{6}{>{\centering\arraybackslash}X}}
				\Xhline{2pt} 
				\multicolumn{1}{l}{\multirow{2}{*}[-0.5ex]{\textbf{Dataset}}} & 
				\multicolumn{1}{l}{\multirow{2}{*}[-0.5ex]{\textbf{Method}}} & 
				\multicolumn{4}{c}{\textbf{Perception}} & 
				\multicolumn{2}{c}{\textbf{Distortion}}  \\
				\cmidrule(lr){3-6} \cmidrule(lr){7-8}
				& & \textbf{DISTS} & \textbf{LPIPS}  & \textbf{KID} & \textbf{FID}  & \textbf{PSNR} & \textbf{MS-SSIM} \\
				\Xhline{1pt} 
				\multirow{11}{*}{HEVC}
				& HM-16.25 \cite{HM} & 0.7 / -0.0002 & 17.6 / -0.0088 & -17.0 / 0.0056 & -11.0 / 2.9019 & 37.8 / -0.9235 & 33.0 / -0.0057 \\
				& VTM-17.0 \cite{VTM} & 0.0 / 0.0000 & 0.0 / 0.0000 & 0.0 / 0.0000 & 0.0 / 0.0000 & 0.0 / 0.0000 & 0.0 / 0.0000 \\
				& DCVC-TCM \cite{DCVC-TCM} & 105.8 / -0.0249 & 53.1 / -0.0201 & 67.9 / -0.0188 & 61.8 / -13.8688 & 38.7 / -0.9163 & 10.2 / -0.0015 \\
				& DCVC-HEM \cite{DCVC-HEM} & 53.9 / -0.0140 & 9.2 / -0.0040 & 31.0 / -0.0091 & 26.3 / -6.3837 & -0.3 / -0.0097 & -15.3 / 0.0019 \\
				& DCVC-DC \cite{DCVC-DC} & 25.2 / -0.0073 & -7.0 / 0.0029 & 15.4 / -0.0049 & 11.2 / -2.9970 & \textcolor{myblue}{\textbf{-19.8 / 0.5900}} & \textcolor{myblue}{\textbf{-28.5 / 0.0042}} \\
				& DCVC-FM \cite{DCVC-FM} & 21.0 / -0.0063 & -7.3 / 0.0032 & 13.5 / -0.0045 & 8.3 / -2.2913 & -16.4 / 0.4788 & -23.3 / 0.0036 \\
				& SEVC \cite{SEVC} & 25.7 / -0.0078 & -4.9 / 0.0022 & 19.4 / -0.0066 & 14.8 / -4.0647 & \textcolor{myred}{\textbf{-25.6 / 0.8075}} & \textcolor{myred}{\textbf{-32.0 / 0.0052}} \\
				& DVC-P \cite{DVC-P} & 122.9 / -0.0278 & 195.2 / -0.0609 & 75.3 / -0.0167 & 96.7 / -18.3528 & 359.0 / -4.2982 & 241.0 / -0.0211 \\
				& PLVC \cite{PLVC} & -27.1 / 0.0110 & \textcolor{myblue}{\textbf{-66.9 / 0.0518}} & -47.3 / 0.0216 & -28.7 / 10.7095 & 262.4 / -3.7475 & 120.0 / -0.0209 \\
				& DiffVC \cite{DiffVC} & \textcolor{myblue}{\textbf{-79.0 / 0.0391}} & -64.8 / 0.0401 & \textcolor{myblue}{\textbf{-75.8 / 0.0281}} & \textcolor{myblue}{\textbf{-47.8 / 14.0877}} & N/A / -6.4813 & N/A / -0.0453 \\
				& DiffVC-OSD & \textcolor{myred}{\textbf{N/A / 0.0428}} & \textcolor{myred}{\textbf{-77.1 / 0.0457}} & \textcolor{myred}{\textbf{N/A / 0.0285}} & \textcolor{myred}{\textbf{-56.9 / 14.6704}} & N/A / -6.4733 & 657.8 / -0.0457 \\ \Xhline{1pt} 
				\multirow{11}{*}{MCL-JCV}
				& HM-16.25 \cite{HM} & 0.6 / -0.0001 & 23.7 / -0.0118 & -2.6 / 0.0003 & 6.2 / -0.8585 & 40.2 / -0.7545 & 37.5 / -0.0049 \\
				& VTM-17.0 \cite{VTM} & 0.0 / 0.0000 & 0.0 / 0.0000 & 0.0 / 0.0000 & \textcolor{myblue}{\textbf{0.0 / 0.0000}} & 0.0 / 0.0000 & 0.0 / 0.0000 \\
				& DCVC-TCM \cite{DCVC-TCM} & 190.5 / -0.0328 & 124.3 / -0.0378 & 125.7 / -0.0087 & 96.9 / -8.5371 & 24.9 / -0.4232 & 24.5 / -0.0024 \\
				& DCVC-HEM \cite{DCVC-HEM} & 128.7 / -0.0249 & 79.3 / -0.0270 & 93.4 / -0.0069 & 64.9 / -6.1862 & -10.9 / 0.2178 & -1.3 / -0.0000 \\
				& DCVC-DC \cite{DCVC-DC} & 95.8 / -0.0208 & 56.6 / -0.0213 & 64.0 / -0.0053 & 44.9 / -4.7827 & \textcolor{myblue}{\textbf{-22.3 / 0.4831}} & \textcolor{myblue}{\textbf{-10.4 / 0.0011}} \\
				& DCVC-FM \cite{DCVC-FM} & 90.2 / -0.0210 & 58.5 / -0.0224 & 67.8 / -0.0061 & 48.4 / -5.6161 & -16.1 / 0.3549 & -7.0 / 0.0008 \\
				& SEVC \cite{SEVC} & 89.6 / -0.0211 & 56.4 / -0.0217 & 62.9 / -0.0058 & 38.4 / -4.6186 & \textcolor{myred}{\textbf{-28.6 / 0.6769}} & \textcolor{myred}{\textbf{-21.0 / 0.0026}} \\
				& DVC-P \cite{DVC-P} & 160.8 / -0.0334 & 236.5 / -0.0753 & 186.6 / -0.0117 & \mbox{237.2 / -18.1484} & 465.4 / -3.8693 & 388.0 / -0.0214 \\
				& PLVC \cite{PLVC} & -67.6 / 0.0300 & N/A / 0.0857 & \textcolor{myblue}{\textbf{-58.3 / 0.0082}} & 19.5 / -2.0915 & 384.6 / -3.9255 & 296.0 / -0.0247 \\
				& DiffVC \cite{DiffVC} & \textcolor{myblue}{\textbf{N/A / 0.0556}} & \textcolor{myblue}{\textbf{N/A / 0.0868}} & -52.5 / 0.0054 & 0.2 / -0.3386 & N/A / -6.3123 & 509.2 / -0.0296 \\
				& DiffVC-OSD & \textcolor{myred}{\textbf{N/A / 0.0694}} & \textcolor{myred}{\textbf{N/A / 0.1015}} & \textcolor{myred}{\textbf{N/A / 0.0104}} & \textcolor{myred}{\textbf{-60.1 / 5.7259}} & \mbox{1026.2 / -4.9601} & 444.4 / -0.0261 \\ \Xhline{1pt} 
				\multirow{11}{*}{UVG}
				& HM-16.25 \cite{HM} & -16.1 / 0.0048 & 3.6 / -0.0014 & -5.7 / 0.0005 & \textcolor{myblue}{\textbf{-0.2 / -0.0011}} & 36.0 / -0.6329 & 29.2 / -0.0046 \\
				& VTM-17.0 \cite{VTM} & 0.0 / 0.0000 & 0.0 / 0.0000 & 0.0 / 0.0000 & 0.0 / 0.0000 & 0.0 / 0.0000 & 0.0 / 0.0000 \\
				& DCVC-TCM \cite{DCVC-TCM} & 165.7 / -0.0237 & 94.8 / -0.0300 & 126.9 / -0.0036 & 92.1 / -3.1260 & 22.4 / -0.3524 & 27.2 / -0.0032 \\
				& DCVC-HEM \cite{DCVC-HEM} & 112.3 / -0.0178 & 62.1 / -0.0215 & 90.5 / -0.0026 & 55.7 / -1.9832 & -14.0 / 0.2809 & 0.4 / -0.0002 \\
				& DCVC-DC \cite{DCVC-DC} & 89.4 / -0.0160 & 43.5 / -0.0168 & 85.3 / -0.0031 & 54.1 / -2.2883 & \textcolor{myblue}{\textbf{-25.8 / 0.5460}} & \textcolor{myblue}{\textbf{-11.6 / 0.0015}} \\
				& DCVC-FM \cite{DCVC-FM} & 95.1 / -0.0181 & 38.1 / -0.0151 & 65.9 / -0.0030 & 39.9 / -2.0906 & -20.4 / 0.4366 & -8.1 / 0.0011 \\
				& SEVC \cite{SEVC} & 105.8 / -0.0194 & 43.4 / -0.0168 & 51.1 / -0.0022 & 38.1 / -1.8837 & \textcolor{myred}{\textbf{-30.6 / 0.6825}} & \textcolor{myred}{\textbf{-16.3 / 0.0023}} \\
				& DVC-P \cite{DVC-P} & 199.4 / -0.0325 & 226.3 / -0.0728 & 306.0 / -0.0080 & 295.8 / -9.2418 & 528.6 / -4.0096 & 434.7 / -0.0274 \\
				& PLVC \cite{PLVC} & N/A / 0.0336 & \textcolor{myblue}{\textbf{N/A / 0.0873}} & \textcolor{myblue}{\textbf{-66.4 / 0.0045}} & 6.7 / -0.0352 & 568.5 / -4.2150 & 374.9 / -0.0284 \\
				& DiffVC \cite{DiffVC} & \textcolor{myblue}{\textbf{N/A / 0.0464}} & -78.6 / 0.0722 & -27.4 / 0.0020 & 32.6 / -1.3766 & N/A / -5.5030 & 486.3 / -0.0330 \\
				& DiffVC-OSD & \textcolor{myred}{\textbf{N/A / 0.0621}} & \textcolor{myred}{\textbf{N/A / 0.0913}} & \textcolor{myred}{\textbf{N/A / 0.0066}} & \textcolor{myred}{\textbf{-51.3 / 2.1179}} & \mbox{1060.3 / -4.7638} & 498.6 / -0.0316 \\
				\Xhline{2pt}  
			\end{tabularx}
			\begin{tablenotes}
				\item[*] \textbf{\textcolor{myred}{Red}} and \textbf{\textcolor{myblue}{Blue}} indicate the best and the second-best performance, respectively. 'N/A' indicates that BD-rate cannot be calculated due to the lack of overlap. 
			\end{tablenotes}
		\end{threeparttable}
	}
	\vspace{-15pt}
\end{table*}

\subsection{End-to-End Finetuning Strategy} \label{sec:E2EFinetuning}
DiffVC-OSD adopts a nine-stage training strategy to fully optimize each component of the model. The first seven stages follow the same procedures as those used in DiffVC \cite{DiffVC}. In stage eight, to adapt the Denoising Unet to the video compression task, we apply LoRA \cite{LoRA} for lightweight fine-tuning, while freezing all other modules except for the TCA and the LoRA parameters. The model is optimized using the following loss function:
\begin{equation}
	\begin{aligned}
		L_8= & w_1 MSE(x_t,\hat{x}_t)+ \\
		& w_2 LPIPS(x_t,\hat{x}_t)+w_3 DISTS(x_t,\hat{x}_t)
	\end{aligned}
\end{equation}
where $w_1$, $w_2$ and $w_3$ are the weights for distortion or perceptual loss terms. Typically, multi-step diffusion-based NVC methods require iterative inference, which makes end-to-end training difficult due to the high computational overhead associated with gradient accumulation. In contrast, DiffVC-OSD performs only a single diffusion step, enabling efficient end-to-end finetuning without such burdens. Specifically, in stage nine, we optimize the entire DiffVC-OSD framework using a rate-distortion-perception loss defined as:
\begin{equation}
	\begin{aligned}
		L_9= & R(v_t)+R(y_t)+ w_t \cdot \lambda (w_4 MSE(x_t,\hat{x}_t)+ \\
		& w_5 LPIPS(x_t,\hat{x}_t)+w_6 DISTS(x_t,\hat{x}_t))
	\end{aligned}
\end{equation}
where $R(v_t)$ and $R(y_t)$ denote the bitrates of the motion vector $v_t$ and the current frame’s latent representation $y_t$, respectively. The coefficients $w_4$, $w_5$, and $w_6$ are the weights for the respective loss terms, and $\lambda$ controls the trade-off between bitrate and reconstruction quality. Additionally, $w_t$ is a periodically varying weight following DCVC-DC \cite{DCVC-DC}.

\section{Experiments}
\vspace{-5pt}
\subsection{Experimental Setting}
\vspace{-5pt}
\textbf{Training:} We use the Vimeo-90k dataset \cite{Vimeo90k} for training. Following the same setup as DiffVC \cite{DiffVC}, we set four $\lambda$ values: $\{16, 48, 128, 384\}$. In stage eight, the weights $w_1$, $w_2$ and $w_3$ are set to 10.0, 1.0 and 1.0, respectively. The batch size is set to 30, and the model is trained for 7 epochs. The initial learning rate is 3e-4 and is decayed to 1e-4, 5e-5, and 1e-5 at the 3rd, 5th, and 7th epochs, respectively. In stage nine, the weights $w_4$, $w_5$ and $w_6$ are set to 0.8, 0.08 and 0.08, and the $w_t$ for four consecutive frames are set to $\{0.5, 1.2, 0.5, 0.9\}$. The batch size is set to 10, and the model is trained for 1 epoch with a constant learning rate of 5e-6. All experiments are conducted on RTX 3090 GPUs.

\textbf{Evaluation:} We evaluate the proposed method on three benchmark datasets: HEVC \cite{HEVC}, MCL-JCV \cite{MCL-JCV}, and UVG \cite{UVG}. Following prior works \cite{DCVC-DC, DiffVC}, we evaluate the first 96 frames of each video sequence, with the intra-period set to 32. The low-delay encoding configuration is adopted, and I-frames are encoded using DCVC-DC Intra \cite{DCVC-DC}. To comprehensively and quantitatively evaluate compression performance, we adopt several established metrics. For perceptual metrics, we use LPIPS \cite{LPIPS}, DISTS \cite{DISTS}, KID \cite{KID} and FID \cite{FID}. For distortion metrics, we report PSNR and MS-SSIM \cite{MS-SSIM}. Finally, we use Bits Per Pixel (BPP) to measure the bitrate consumption per pixel per frame.

\begin{figure}[t]
	\centering
	\includegraphics[width=0.5\textwidth]{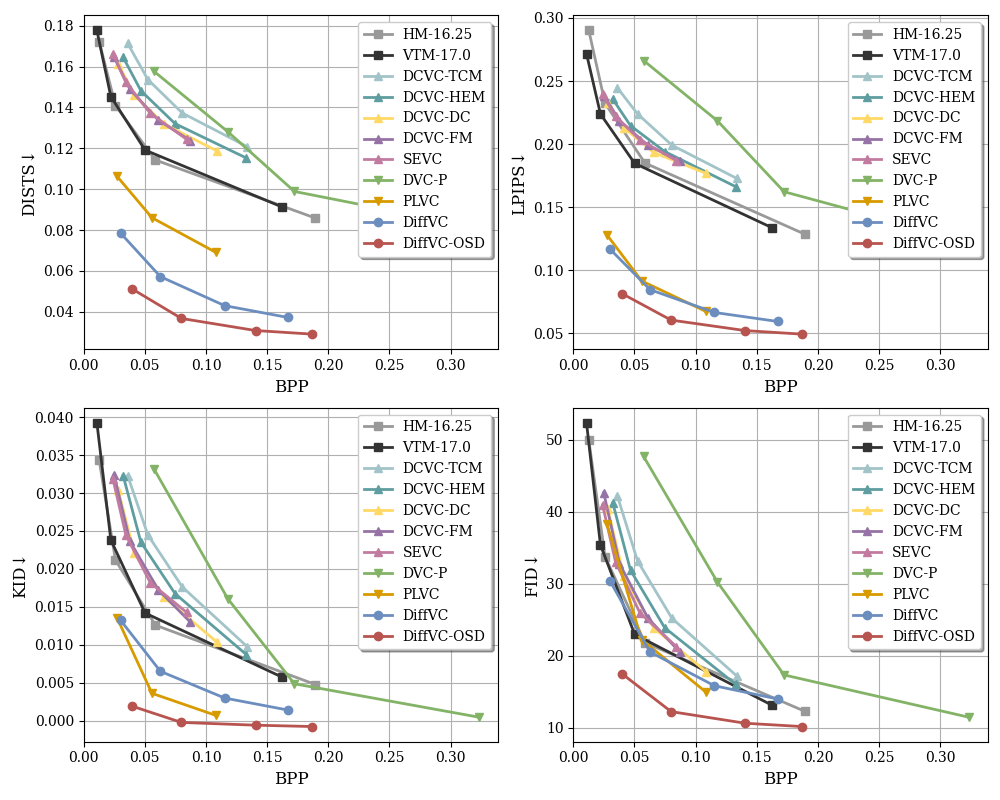}
	\caption{The rate-perception curves of the DiffVC-OSD and other video compression methods on the MCL-JCV dataset.}
	\label{fig:MCL-JCV_RD_Curve}
	\vspace{-5pt}
\end{figure}

\begin{figure}[t]
	\centering
	\includegraphics[width=0.5\textwidth]{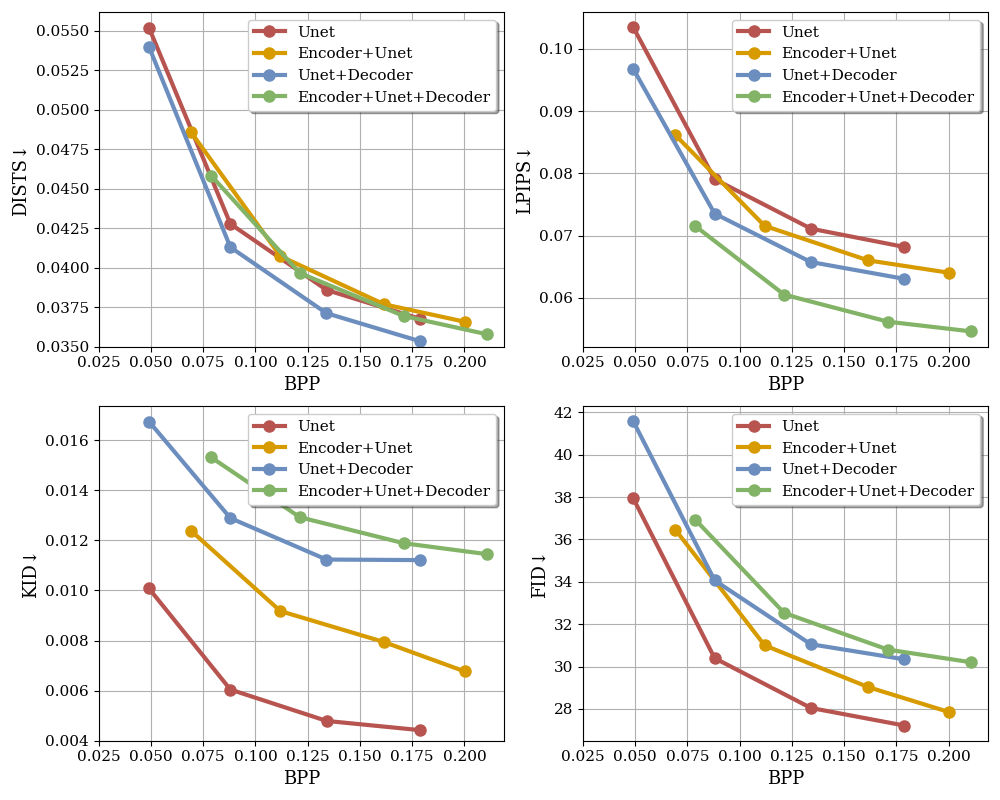}
	\caption{The ablation of lora position.}
	\label{fig:lora_position}
	\vspace{-15pt}
\end{figure}

\subsection{Experimental Results and Analysis}
To evaluate different categories of video compression methods, we select several representative approaches from each category for comparison. Traditional codecs include HM-16.25 \cite{HM} and VTM-17.0 \cite{VTM}. Distortion-oriented NVCs include DCVC-TCM \cite{DCVC-TCM}, DCVC-HEM \cite{DCVC-HEM}, DCVC-DC \cite{DCVC-DC}, DCVC-FM \cite{DCVC-FM}, and SEVC \cite{SEVC}. GAN-based NVCs include DVC-P \cite{DVC-P} and PLVC \cite{PLVC}, while diffusion-based NVCs include DiffVC \cite{DiffVC} and our proposed DiffVC-OSD.
Figure \ref{fig:MCL-JCV_RD_Curve} shows the rate–perception curves of these methods on the MCL-JCV dataset. Table \ref{tab:results} summarizes their compression performance across three benchmark datasets, measured using BD-rate and BD-metric, with VTM-17.0 as the anchor. It is important to note that a lower BD-rate and a higher BD-metric indicate better compression performance.
Experimental results demonstrate that DiffVC-OSD achieves state-of-the-art performance on all perceptual metrics across all test datasets. Even with respect to distortion metrics, DiffVC-OSD outperforms the multi-step diffusion-based method DiffVC.

\subsection{Ablation Studies}
Table \ref{tab:ablation} presents the results of our ablation studies. MSD, OSD, TCA, and FT refer to the Multi-Step Diffusion Model, One-Step Diffusion Model, Temporal Context Adapter, and End-to-End Finetuning strategy, respectively. Notably, in this ablation study, MSD employs 50 diffusion steps, while all other settings remain consistent with OSD. The comparison between Method A and Method B shows that OSD achieves a significant improvement in perceptual quality by leveraging its strong generative prior, with only a 0.2s increase in decoding time. Comparing Method B and Method C, we observe that incorporating TCA yields an average improvement of 10.61\% in perceptual compression performance, with a negligible decoding time increase of 0.01s. The comparison between Method C and DiffVC-OSD shows that the End-to-End Finetuning strategy results in an average 4.11\% bitrate reduction on perceptual metrics. Finally, the comparison between Method C and Method D confirms that, compared to MSD, OSD provides an average 82.81\% improvement in perceptual compression performance and achieves nearly 20$\times$ faster decoding speed.

\begin{figure}[t]
	\centering
	\includegraphics[width=0.5\textwidth]{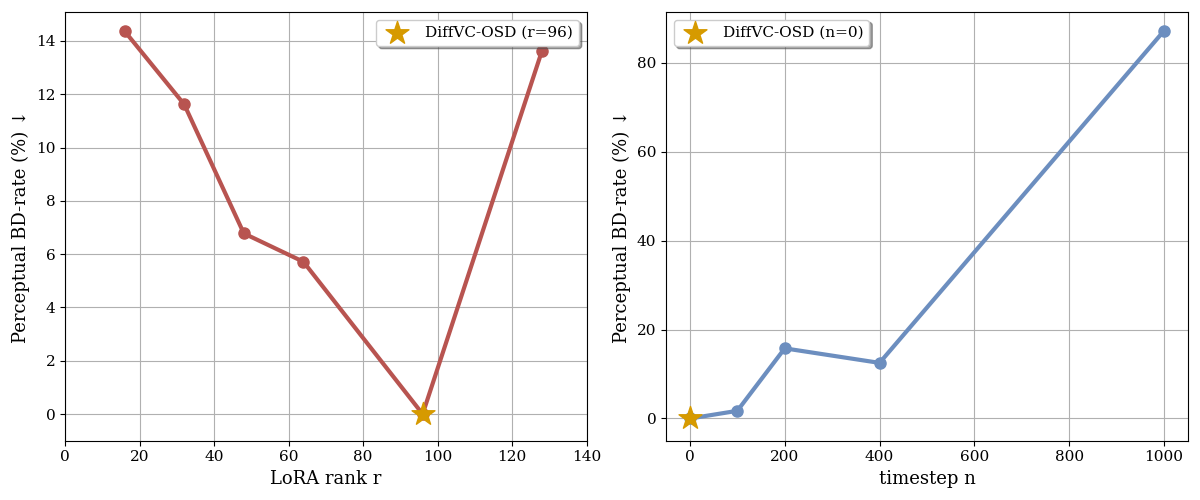}
	\caption{The ablation of LoRA rank $r$ and timestep $n$.}
	\label{fig:lora_rank_and_timestep}
	\vspace{-15pt}
\end{figure}

Additionally, as illustrated in Figure \ref{fig:lora_position}, we examine the impact of inserting LoRA at different locations within the Stable Diffusion architecture. To achieve a balanced performance across four perceptual metrics, we choose to insert LoRA only into the Unet. Furthermore, Figure \ref{fig:lora_rank_and_timestep} explores how varying the LoRA rank $r$ and timestep $n$ fed into the Unet affects performance. Based on the results, we adopt $r=96$ and $n=0$ as the default configuration.

\begin{table}[t]
	\centering
	\scriptsize
	\renewcommand {\arraystretch}{1.1}
	\caption{The ablation of DiffVC-OSD. All results are tested on HEVC Class C with DiffVC-OSD as the anchor.}
	\label{tab:ablation}
	\resizebox{\linewidth}{!}{
		\begin{threeparttable}
			\begin{tabularx}{\linewidth}{@{\hspace{2.5pt}}c@{\hspace{2.5pt}} *{4}{c@{\hspace{2.5pt}}} @{\hspace{3pt}}  *{5}{c@{\hspace{3pt}}} c@{\hspace{3pt}}}
				\Xhline{1.2pt}
				\multicolumn{1}{c}{\multirow{2}{*}[-0.5ex]{\textbf{Method}}} & 
				\multicolumn{4}{c}{\textbf{Modules}} & 
				\multicolumn{5}{c}{\textbf{BD-rate (\%) $\downarrow$}} & 
				\multicolumn{1}{c}{\multirow{2}{*}[-0.5ex]{\shortstack{\textbf{Dec.} \\ \textbf{Time (s) $\downarrow$}}}} \\
				\cmidrule(lr){2-5} \cmidrule(lr){6-10}
				& \textbf{MSD} & \textbf{OSD} & \textbf{TCA} & \textbf{FT} 
				& \textbf{DISTS} & \textbf{LPIPS} & \textbf{KID} & \textbf{FID} & \textbf{Avg.} 
				& \\
				\Xhline{0.8pt}
				A & \ding{55} & \ding{55} & \ding{55} & \ding{55} & 102.38 & 192.37 & N/A & 234.31 & N/A & 0.19 \\
				B & \ding{55} & \checkmark & \ding{55} & \ding{55} & 6.99 & 7.36 & 34.02 & 10.50 & 14.72 & 0.39 \\
				C & \ding{55} & \checkmark & \checkmark & \ding{55} & 4.77 & 5.35 & 4.56 & 1.75 & 4.11 & 0.40 \\
				D & \checkmark & \ding{55} & \checkmark & \ding{55} & 56.92 & 61.45 & 156.41 & 72.91 & 86.92 & 8.14 \\
				DiffVC-OSD & \ding{55} & \checkmark & \checkmark & \checkmark & 0.00 & 0.00 & 0.00 & 0.00 & 0.00 & 0.40 \\
				\Xhline{1.2pt}
			\end{tabularx}
			\begin{tablenotes}
				\item[*] Dec. Time refers to the time required to decode a P frame. 'N/A' indicates that BD-rate cannot be calculated due to the lack of overlap.
			\end{tablenotes}
		\end{threeparttable}
	}
\end{table}

\section{Conclusions}
In this paper, we introduced DiffVC-OSD, a novel One-Step Diffusion-based Perceptual Neural Video Compression framework. By directly utilizing noise-free reconstructed latent representations and applying a single-step diffusion process guided by temporal context, DiffVC-OSD achieves high perceptual quality with significantly lower decoding time. The proposed Temporal Context Adapter enhances conditional guidance, while End-to-End Finetuning further boosts overall compression efficiency. Extensive experiments demonstrate that DiffVC-OSD achieves state-of-the-art perceptual compression performance, delivering a 86.92\% bitrate reduction and about 20$\times$ faster decoding compared to its multi-step diffusion-based counterpart. Our work highlights the potential of onse-step diffusion models in advancing perceptual video compression.

\bibliography{reference}
\end{document}